\documentclass[amsmath,amssymb,prl,aps,twocolumn,mathbbm,superscriptaddress]{revtex4-1}
\usepackage{graphicx,dsfont}
\usepackage{epsfig}
\usepackage{physics}
\usepackage[normalem]{ulem}
\usepackage[dvipsnames]{xcolor}
\usepackage[colorlinks=true,citecolor=Cerulean,linkcolor=RubineRed,urlcolor=Cerulean]{hyperref}

\begin{document}

\title{Decoherence-Free Rotational Degrees of Freedom for Quantum Applications}

\author{J. S. Pedernales, F. Cosco, and M. B. Plenio}
\affiliation{Institut f\"ur Theoretische Physik und IQST, Albert-Einstein-Allee 11, Universit\"at Ulm, D-89081 Ulm, Germany}

\begin{abstract}

We employ spherical $t$-designs for the systematic construction of solids whose rotational degrees of freedom can be
made robust to decoherence due to external fluctuating fields while simultaneously retaining their sensitivity to 
signals of interest. Specifically, the ratio of signal phase accumulation rate from a nearby source to the 
decoherence rate caused by fluctuating fields from more distant sources can be incremented to any desired level by using
increasingly complex shapes. This allows for the generation of long-lived macroscopic quantum superpositions of
rotational degrees of freedom and the robust generation of entanglement between two or more such solids with applications
in robust quantum sensing and precision metrology as well as quantum registers.

\end{abstract}

\maketitle

{\it Introduction.---} Quantum metrology and sensing represent promising near term applications 
of quantum technologies as they require the control of a few
or even single quantum
systems. A wide variety of physical systems are being developed for these purposes ranging from 
solid state spins \cite{WuJP+16} to ultracold trapped atoms and ions \cite{LudlowBY+15}, atom 
interferometry \cite{CroninSP09}, optomechanical systems \cite{AspelmeyerKM14} and levitated
massive particles \cite{LibbrechtB04}.

A central challenge common to all quantum sensor designs is the necessity to achieve both, robustness 
to environmental noise and, at the same time, high sensitivity to a signal of interest.
To this end, while affecting the signal as little as possible, we need to either filter noise, correct 
for its effect or encode the system to reduce its sensitivity to environmental noise. This becomes possible
whenever signal and noise have some characteristic in which they differ. Frequently used examples include 
cases in which signal and noise differ in spectral or temporal support, in the symmetries of signal and 
noise or in the operator with which they act on the system. These allow us to separate signal and noise, e.g.,
via dynamical decoupling \cite {Viola1999,Ryan2010,Lange2010,Naydenov2012,Cai2013,Mueller2014,Casanova2015}, 
time-gating, decoherence-free sub-spaces \cite{PalmaSE96,PlenioVK97} or quantum error correction 
\cite{UndenBL+16}, respectively. 

The principles of levitation and feedback cooling of massive particles in high vacuum have been demonstrated by Ashkin \cite{AshkinD1976a,AshkinD1976b} and with it came the recognition that such systems are promising for ultraprecise metrology and sensing due to the lack of friction and a high degree of isolation from nearby sources of noise and decoherence. Reference \cite{LibbrechtB04} was the first to propose their use in the quantum regime and this has led to theoretical and experimental work towards cooling of optically levitated silica 
nano-microspheres \cite{Raizen2010,BarkerS2010,Zoller2010,Raizen2011,Cirac2011,Delic2019,Delic2020}, magnetically 
levitated diamond \cite{Hsu2016} and silica nanocrystals \cite{Hsu2018}. The control of single electron spins
in trapped diamond nanocrystals was first demonstrated in \cite{GeiselmannJR+2013,Vamivakas2013,Vamivakas2017} 
motivating subsequent theoretical explorations \cite{Bose2013,Duan2013,Plenio2014} as well as further experimental 
work that include the control of their torsional degrees of freedom under optical forces \cite{Li2016b} and ion traps 
\cite{DelordNC+2017,DelordHN+2019}. 
An important long term goal in this field is the realisation of tests for possible minute deviations from quantum 
physics \cite{Bouwmeester2003,Romero-Isart2011,Bose2013,Duan2013,Plenio2014,BahramiPB+14}, for the detection of 
corrections of known force laws due to extra dimensions \cite{LiuZ19} and new forces and 
particles \cite{CarneyGK+19,CarneyHL+19} or the realisation of early proposals of gravitationally induced entanglement 
\cite{Feynman1957,KafriTM14} using levitated particles \cite{BahramiBM+15,SchmoeleDHA16,Bose2017,Krisnanda2019}. 
A common goal of many of these approaches is the detection of minute interactions with nearby sources 
in the presence of noise from more distant sources. However, current proposals for matter-wave interferometry 
with translational degrees of freedom of massive particles are susceptible to first order (gradient) fluctuations 
and hence perturbing signals of the same nature emanating even from distant sources~\cite{Pedernales2019}. Therefore, 
we are faced with the challenge of suppressing perturbations from distant sources without suppressing the desired 
signals from nearby sources. 

In  this  Letter,  we  address this challenge by designing the shape of rigid bodies such that their {\em rotational} degrees of freedom can be made robust against 
decoherence from distant sources, while at the same time allowing for interaction with signals from nearby
sources. To this end we introduce a systematic method, based on the mathematical theory of spherical $t$-designs, 
to construct rigid bodies whose rotational states are degenerate up to a desired order of the multipole 
expansion of their energy in a perturbing potential. In this manner, we ensure that superpositions in
the basis of rotational states of a suitably designed solid can, in principle, exhibit arbitrarily long 
coherence times in the presence of perturbations from distant sources. Moreover, we show that the ratio 
of the entangling to decoherence rate between two of these objects scales favourably with the order of 
the spherical $t$-design.

{\it Multipole expansion and spherical $t$-designs.---} 
Consider a rigid body that is held in free space and whose rotational degrees of freedom we aim to control 
and place in superposition of different orientations of this body. The elimination of decoherence of such a superposition requires the construction of an object whose potential energy in an 
external field is the same for all of its orientations, so that when the object is placed in a superposition of 
two different orientations, fluctuations of the perturbing field translate into global phase variations, 
keeping the relative phase of the superposition constant, and in this manner preserving its coherence. A shape 
which, thanks to its rotational symmetry, trivially fulfils such a goal is that of a uniform sphere. However, 
it is important to note that such an object is, for our purposes, useless as the different orientations of a sphere
are indistinguishable, and, as a consequence, superpositions of different orientations of the sphere become
experimentally inaccessible. Therefore, our goal will be the design of a body that decouples from the external 
fields as well as possible and, simultaneously, deviates as much as possible from the spherical symmetry, such
that different orientations of such a body can be placed in superposition and can be made to interact between 
two such bodies.

For a system whose center of mass is placed at the origin of coordinates, the perturbing field generated 
by a distant source admits a multipole expansion
\begin{equation}
\label{multipoleexp}
    \phi(\vec x) = \sum_{n=0}^{\infty} \sum_{i_1,\ldots,i_n = 1}^3 M_{i_1,\ldots,i_n} (\vec x)_{i_1}\cdot\ldots\cdot (\vec x)_{i_n},
\end{equation}
where $M_{i_1,\ldots,i_n} $ are the $n$-th order moments of the field. The $n$-th order moments will scale as 
$1/L^{n}$, where $L$ is the distance to the object generating the perturbing field. The total energy of an object 
with a density distribution $\rho(\vec x)$ in such a potential is given by $$V=\int d\vec {x} \rho(\vec {x}) \phi(\vec {x}).$$
Hence, for perturbing fields that originate at distances much bigger than the spatial extent $R$ of the body, the
contribution of the $n$-th order moment to the energy will scale as $(R/L)^n$ and consequently become rapidly negligible. 

In order to reduce decoherence from distant fields, we will engineer a density distribution that results in 
rotationally invariant contributions to $V$ up to a given order of the multipole expansion so that only
higher moments can contribute to decoherence. For simplicity, 
we construct our distribution as $\rho (\vec x)=q \sum_{i=1}^N  \delta (\vec x - \vec P_i)$, that is, a 
collection of a finite number $N$ of point charges/masses $q$ located at positions $\vec P_i$. A systematic 
strategy to achieve this makes use of the concept of spherical $t$-designs \cite{Bajnok1992} which is a set 
of $N$ points $\{ \vec P_i \}$ on the unit sphere of dimension $d+1$, ${ S^d = \left\{\vec x \in 
\mathbb{R}^{d+1}: \vec x \cdot \vec x = 1\right\} }$, which satisfies
\begin{equation}
    \int_{S^d} f(\vec x)d\mu(\vec x) = \frac{1}{N}\sum_{i=1}^N f(\vec P_i) \label{tdesign},
\end{equation}
for all polynomials of degree less or equal to $t$, ${f(\vec x) = \sum_{n=1}^{t} \sum_{i_1,\ldots,i_n = 1}^3 c_{i_1,\ldots,i_n} (\vec x)_{i_1}\ldots (\vec x)_{i_n}}$. 
Equation~(\ref{tdesign}) implies that, for such a density, the contributions to the total energy are invariant under
rigid rotations up to order $t$ in the multipole expansion. This follows from the linearity of the rotation 
operation in terms of coordinate variables, which guarantees that the potential function truncated at order 
$t$ transforms into a polynomial of the same order in a rotated frame. Therefore, up to order $t$, the potential
energy of the spherical $t$-design in the rotated field equals that of a uniform sphere and, as the potential
energy of the sphere is evidently invariant under rotations, so is the energy of the $t$-design up to order 
$t$. Now, to relate this result to our aim of designing a solid with a continuous density distribution whose potential 
energy is invariant under rotations up to order $t$ in the multipole expansion, we note that, for three spatial 
dimensions, any choice of points on the unit sphere can be extended to a simply connected solid by considering the 
set of points on all spheres of all radii $r\le 1$ and adopting a finite volume when allowed to trace out a solid angle. 
For an object constructed in such a manner from a spherical $t$-design, all contributions up to order $t$ in the 
multipole expansion of the potential energy are rotationally invariant. A fluctuating external potential field will induce variations in the relative phase between two
orientations of such an object only in the $(t+1)$-th order of the multipole expansion, whose magnitude decreases 
exponentially with $t$, leading to a strong reduction in the decoherence of its rotational degrees of freedom.
\begin{figure}[t!]
   \centering
   \includegraphics[width=\columnwidth]{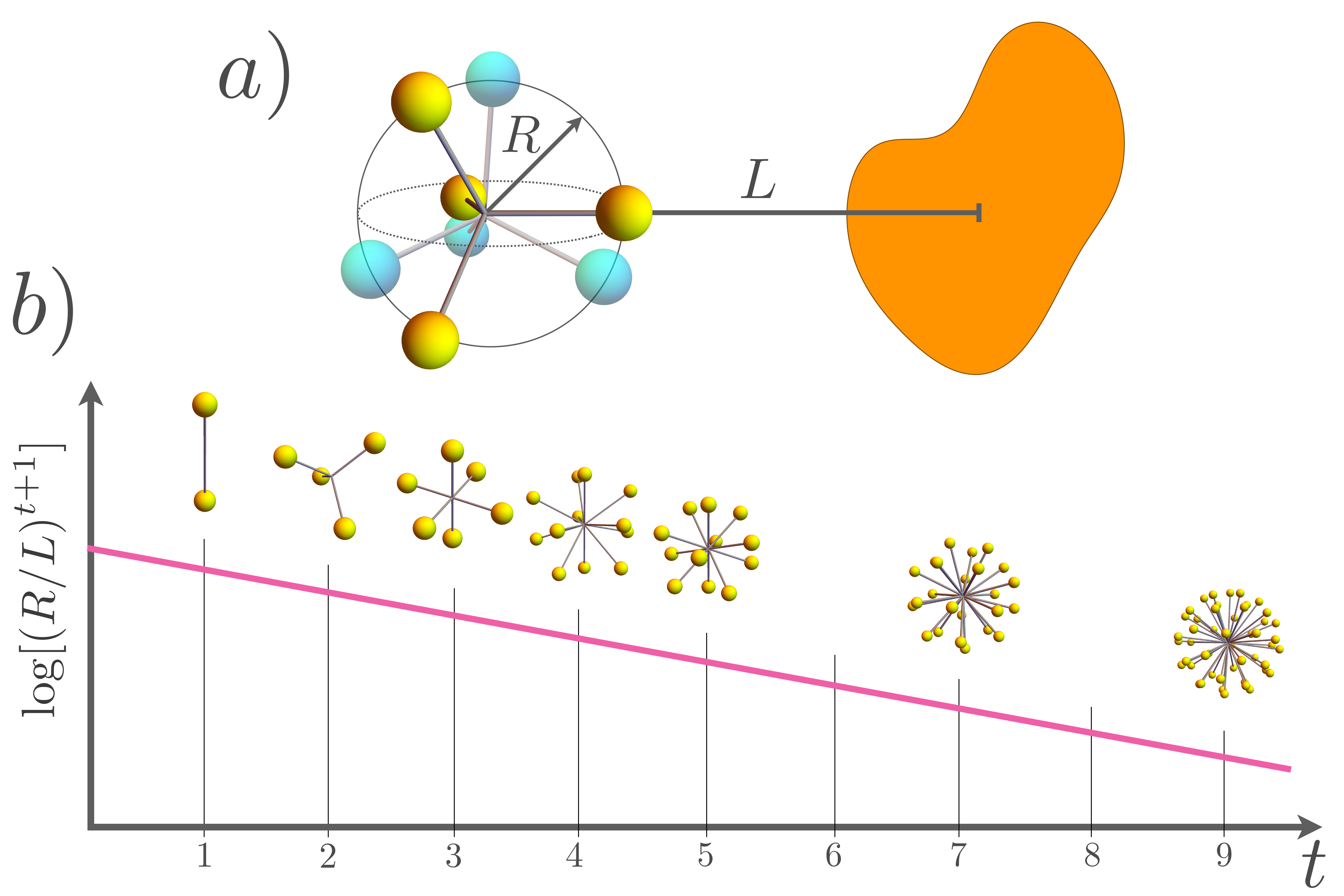} 
   \caption{{\bf $t$-designs in three-dimensional space.}  a) depicts a spherical 2-design consisting of points on a sphere of radius 
   $R$, susceptible to the potential field generated by a distant density distribution (orange object) located at a distance $L$. Two 
   distinct orientations (blue and yellow) of such a spherical $t$-design may have different potential energy, however, this difference 
   will originate from moments of order $3$ and higher in the multipole expansion of the potential field about the center of mass of the 
   body. b) shows analytical solutions of spherical $t$-designs up to order 9 and the scaling of the energy variations that rotations can 
   induce. The leading order in the potential energy difference between two orientations of a $t$-design in the field generated by a distant 
   object is provided by moments of order $t+1$, which for the case of a Coulombian potential, scale following the power law $(R/L)^{t+1}$. }
   \label{fig:t-designs}
\end{figure}

The existence of spherical $t$-designs for any choice of $t$ is well established. In fact, it has been proven that, for each 
$N\ge c_d t^d$, there exists a spherical $t$-design in the sphere $S^d$ consisting of $N$ points, where $c_d$ is a constant 
depending only on $d$ \cite{Bondarenko2013} and, most relevant for our work, it is conjectured that $c_2=1/2$. For orders 
$t=\{1,2,3,4,5,7,9\}$, analytic solutions are known with a number of points $N$ that is close to the optimal number
of elements~\cite{Hardin1996}, while close to optimal numerical solutions have been found up to orders exceeding $t=300$. 

In Fig.~(\ref{fig:t-designs})b we plot known analytic solutions that are conjectured to be optimal in $N$ and leave
out cases where solutions are only numerical. We show for each of these designs the magnitude of the potential energy 
contribution due to the first moment of the perturbing potential that is not rotational invariant, which exhibits an 
exponential decay of the form $(R/L)^{t+1}$; this indicates the magnitude of the fluctuations of the relative phase 
between any two orientations of that object in a perturbing field.

{\it Applications.---}  Objects as described in the previous section may be constructed from nanoparticles and trapped 
either in optical~\cite{Millen2019,Rider2019}, Paul~\cite{Huillery2019} or magneto-gravitational traps~\cite{Hsu2016}. The
rotational degrees of freedom of irregularly shaped nanoparticles have been manipulated in such settings making use of the
torque exerted by circularly polarized light or by magnetic field gradients acting on spin degrees of freedom that the 
nanoparticles may host, e.g. NV-centers in diamond \cite{DelordNC+2017,DelordHN+2019}. In matter-wave interferometry with 
translational degrees of freedom of nanoparticles, superpositions are susceptible to first order (gradient) fluctuations 
of the perturbing fields which, in conjunction with the required long experimental times, set daunting demands on the degree
of isolation of the experimental setups~\cite{Pedernales2019}. Hence, we propose the use of rotational degrees of freedom
of bodies constructed according to spherical $t$-designs which, due to their extended coherence, are expected to result in 
a superior performance. For example, a superposition of two orientations of such a nanoparticle would constitute a macroscopic
quantum superposition whose lifetime can be used to set bounds on free parameters of CSL models \cite{Bassi2013}.

\begin{figure}[t!]
   \centering
   \includegraphics[width=\columnwidth]{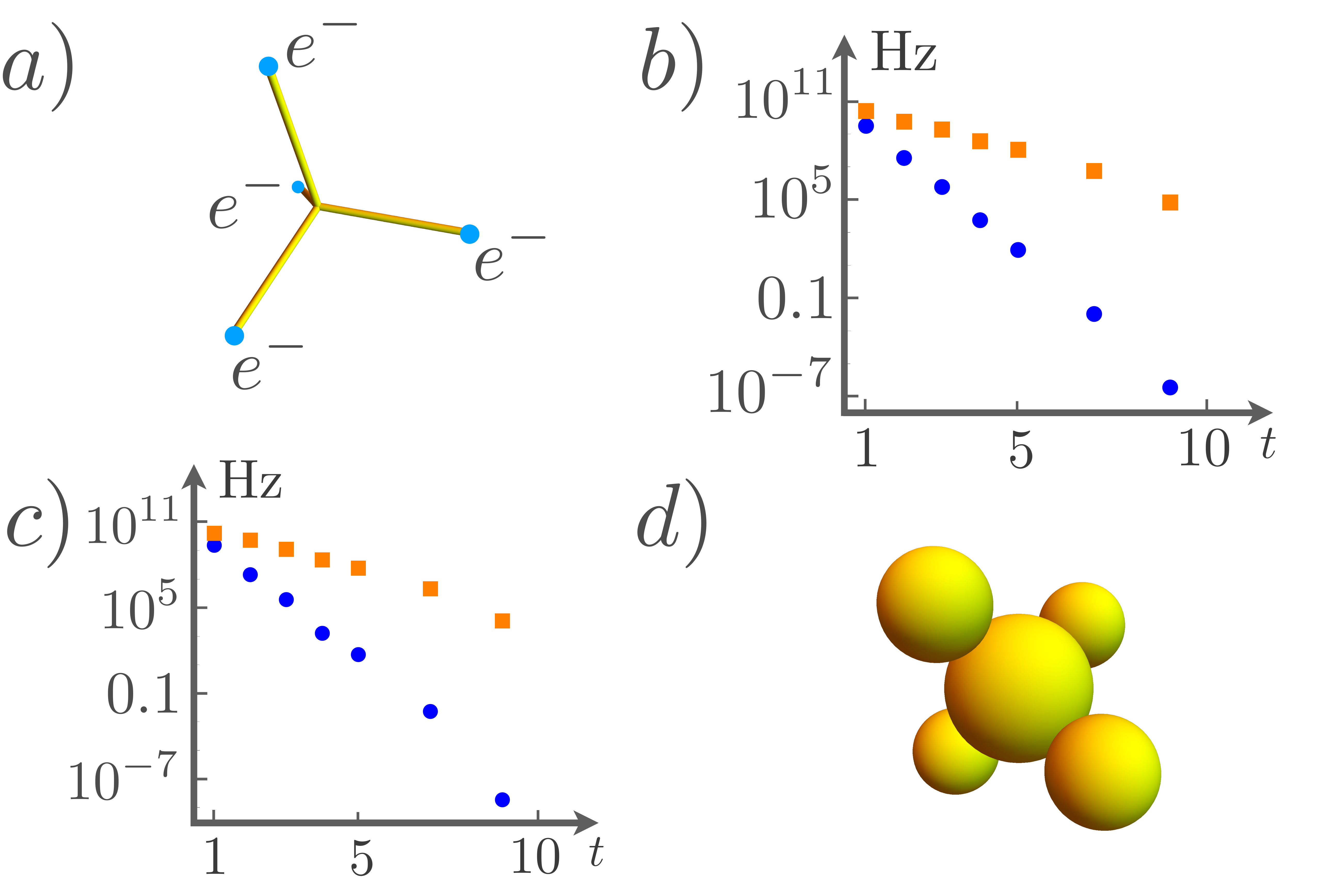}
   \caption{{\bf Numerical analysis.}  a) depicts a possible realisation of a 2-design for electric charges. b) shows 
   the resulting scaling of $\bar \Delta_{\rm signal}$ (orange squares) and $\bar \Delta_{\rm noise}$ (blue dots), as a function of the 
   order of the $t$-designs. c) shows the scaling of magnitudes $E_{\rm ent}$ (orange squares) and $\bar \Delta_{\rm noise}$ (blue dots) 
   for two electrically interacting bodies at a distance of $10 \ \mu$m, each corresponding to a spherical $t$-design with 
   a radius of $2\ \mu$m. In b) and c), the noise source is taken to be a body with total charge $10^3$e$^{-}$ at a distance of $200\ \mu$m, and 
   rotational states $R1$ and $R2$ are numerically optimized to maximize the signal sensitivity and the entangling rate respectively. d) 
   depicts a possible realisation of a $t$-design for the generation of gravitationally mediated entanglement.}
\label{fig:-applications}
\end{figure}

Furthermore, we stress that the enhancement in the isolation of the rotational degrees freedom of $t$-designed objects from 
distant perturbation sources does not prevent them from retaining sensitivity to fields generated by closer sources and thus 
allows for their application as robust quantum sensors. Consider a $t$-designed object that is a distance $L$ from a perturbing 
source and a distance $D$ from the signal source of interest, such that $L/D \gg 1$. When initially prepared in a superposition 
of two orientations $\ket{R1} + \ket{R2}$ this state will acquire a relative phase at a rate $\bar \Delta = \Delta (R1)  - \Delta (R2)$ due 
to the action of signal and perturbing field. As the leading order contributions to this rate due to signal and perturbation, $\bar \Delta 
= \bar \Delta_{\rm signal}  + \bar \Delta_{\rm noise}$, originate from the $(t+1)$-st order contribution in their multipole 
expansion, they scale as $\bar \Delta_{\rm signal}\sim(R/D)^{t+1}$ and $\bar \Delta_{\rm noise}\sim(R/L)^{t+1}$ and their 
ratio as $\bar\Delta_{\rm signal}/\bar\Delta_{\rm noise} \sim (L/D)^{t+1}$. As a consequence, the signal to noise ratio can
be improved by increasing the order of the design $t$. The precise value of the ratio $\bar\Delta_{\rm signal}/\bar\Delta_{\rm noise}$ will, in general, depend on the chosen orientations 
$R1$ and $R2$, and the specific form of the signal and noise sources. As an illustrative example we consider the case of a $t$-designed 
object constructed of electric elementary charges distributed at $2 \ \mu$m distance from the origin at positions determined by the known analytic 
solutions for $t$-designs. In practice, these charges could be placed at the tips of neutral cylinders that meet 
at the center of the sphere and are arranged such that they preserve the symmetry of the $t$-design, see Fig.~{{(\ref{fig:-applications})}a } 
for the case of $t=2$. The source of the perturbation is taken to be a body of $10^3$ elementary charges placed at a distance of $L = 200\  \mu$m. 
The signal is generated by a single charge placed at a distance of $D = 10 \ \mu$m. In Fig.~{(\ref{fig:-applications})b} we show the scaling of
both $\bar \Delta_{\rm noise}$ and $\bar \Delta_{\rm signal}$ with the order to the spherical $t$-design, which clearly evidences that 
the sensitivity of the device improves with the order of the $t$-design.

Following a similar reasoning, one can now consider the possibility of making the source of this signal a second $t$-designed 
object, which can also be placed in a superposition of two distinct orientations. The sensitivity to the signal of this source in 
superposition translates into the capability of generating entanglement between the rotational degrees of freedom of the two objects. 
More precisely, let us consider that two $t$-designed objects, A and B, are placed some distance from each other that ensures 
that their interaction is not negligible, and that they are initially prepared in the product state $(\ket{R1}+\ket{R2})_A\otimes
(\ket{R1}+\ket{R2})_B$. Then their global state after a time $T$ is given by ($\hbar = 1$)
\begin{equation}
\begin{split}
\label{fullstate}
e^{-i (E_{11}+ \Delta_{11})T} \ket{R1\ R1}_{AB} + e^{-i (E_{12} + \Delta_{12})T } \ket{R1\ R2}_{AB} \\
+ e^{-i (E_{21} + \Delta_{21})T }\ket{R2\ R1}_{AB} + e^{-i (E_{22} + \Delta_{22})T }\ket{R2\ R2}_{AB},
\end{split}
\end{equation}
where $E_{ij}$ is the interaction energy between the bodies when object A is in the rotational state $Ri$ and body B in the rotational 
state $Rj$. The state in Eq.~(\ref{fullstate}) will be entangled if $E_{\rm ent}  = [ (E_{11} - E_{12}) + (E_{22} - E_{21}) ] \neq \frac{2\pi}{T} n$ 
for $ n= 0, 1 ...$ and if, at the same time $\bar \Delta T \ll 1$ for both solids, so that the relative phase due to the interaction between 
the two particles is not washed out by the fluctuations of the perturbing field. The latter condition can always be ensured by choosing a 
sufficiently high order in the $t$-design combined with a sufficiently long duration of the experiment.

We provide a numerical analysis of the scaling of the quantities $E_{\rm ent}$ and $\bar \Delta$ with the order of the $t$-design by simulating the case of two $t$-designed objects as the one discussed in Fig.~(\ref{fig:-applications})b with their centers placed at a 
distance of $10 \ \mu$m. In Fig.~{(\ref{fig:-applications})}c we show the scaling of both $E_{\rm ent}$ and $\bar \Delta$, for $t$-designs 
with known analytic solution to ensure numerical accuracy of the entangling and decoherence rates and find that the perturbing field is 
suppressed much faster than $E_{\rm ent}$. This shows that two such bodies can get robustly entangled, suggesting applications of the 
introduced $t$-designed objects in quantum technologies.

One could also envision a more ambitious experiment where the entangling force acting on the two $t$-designed bodies is due to gravity 
alone. We consider massive solids with the properties of $t$-designs as the one shown in Fig.~(\ref{fig:-applications})c, where a central 
solid sphere is used to connect several smaller peripheral ones whose centers of mass are placed at the positions determined by specific 
$t$-design solutions (in the example a $2$-design). Notice that the central sphere does not contribute to energy differences due to rotations, 
as its mass distribution is unchanged under rotations. For the numerical analysis, we take the case of diamond, with a central sphere of 
radius $10 \ \mu$m. We fix the total mass of the object to be $1.83 \cdot 10^{-11}$ Kg, and correspondingly adapt the radius of the peripheral 
spheres for each $t$-design. We place two such objects at a distance of $200 \ \mu$m from each other, such that Casimir-Polder forces do
not exceed the gravitational interaction, and we assume that the objects are free of spin or charge impurities so that it can be considered 
that their interaction is dominated by gravity. For such a configuration, we find that $E_{\rm ent}$ ranges from $\sim 16$ Hz for the $1$-design, 
to $\sim 0.001$ Hz for the $3$-design, while the magnitude of the fluctuating phase $\bar \Delta $ quickly decays from $\sim 7$ Hz for the 
$1$-design to $\sim 10^{-12}$ Hz for the $3$-design when the source of perturbation is taken to be a $100$ Kg mass placed
$20$ m away. The presented results indicate that gravitationally mediated entanglement should be observable in the order of seconds for the first three 
$t$-designs, provided that each object can be put in a superposition of two orthogonal rotational states within their coherence time. 

Finally, we would like to remark that the spherical $t$-designs can also find application in designing decoherence-free
subspaces for spin degrees of freedom. The key idea is to use the geometry of $t$-designs to place the spins in space. 
In particular, consider a set of $N$ spins located at the positions of a specific $t$-design, all of which are in the 
state up, $\ket{1}$, and a second set of $N$ spins prepared in state down, $\ket{0}$, located in the positions of the 
same $t$-design but rotated with respect to the first set. Given that rotations do not affect the value of the energy 
up to order $t$, both sets of spins will see the same energy but with opposite sign, as they are in opposite spin states. 
It becomes evident that the global spin state $\ket {111...000}$ will not acquire any phase up to the order $t$ of a 
perturbing magnetic field, and therefore, that the set $\{ \ket {111...000}, \ket {000...111}\}$ constitutes a robust 
basis against magnetic field fluctuations up to order $t$.

{\it Conclusions.---} We have demonstrated that macroscopic superpositions of different orientations of a solid object whose shape 
is suitably determined following the theory of spherical $t$-designs can be made robust to decoherence due to perturbing 
potential fields from external sources to any desired level. Moreover, we argue theoretically and demonstrate numerically that the ratio between phase accumulated from a signal of interest 
originating from a distance $L_{\rm sig}$ and the rate of decoherence imparted by a field originating from a distance $L_{\rm dec}$ 
scales as $(L_{\rm dec}/L_{\rm sig})^{t+1}$ for $L_{\rm sig}\ll L_{\rm dec}$ and can thus be made arbitrarily large. This suggests
a route for enhancing the sensing capabilities of levitated particles by the replacement of translational with 
rotational degrees of freedom~\cite {Stickler2015,Stickler2018,Stickler2018b} and offers a plethora of applications 
in the realm of quantum technologies ranging from sensing to the systematic exploration of rotational 
degrees of freedom for quantum applications.

{\it Acknowledgments.---} We acknowledge support by the ERC Synergy grant BioQ (Grant No. 319130), the EU projects 
HYPERDIAMOND (Grant No. 667192) and AsteriQs (Grant No. 820394), the QuantERA project NanoSpin, the BMBF project 
DiaPol, the state of Baden-W\"urttemberg through bwHPC, the German Research Foundation (DFG) through Grant No. INST 
40/467-1 FUGG, and the Alexander von Humboldt Foundation through a postdoctoral fellowship.

\end{document}